\documentclass[12pt,preprint]{aastex}

\newcommand{\myemail}{dmd@nju.edu.cn}

\begin{document}
\title{Formation Heights of Extreme Ultraviolet Lines in an Active Region Derived by Correlation of Doppler Velocity
\\ and Magnetic Field}
\author{Y. Guo, M. D. Ding, and M. Jin}
\affil{Department of Astronomy, Nanjing University, Nanjing
210093, China}
\email{\myemail}

\and

\author{T. Wiegelmann}
\affil{Max-Planck-Institut f\"{u}r Sonnensystemforschung,
Max-Planck-Strasse 2, 37191 Katlenburg-Lindau, Germany}
\email{wiegelmann@mps.mpg.de}

\begin{abstract}

We study the correlation heights, which indicate the formation
height of Extreme Ultraviolet (EUV) lines in an active region using
observations from the EUV Imaging Spectrometer (EIS) and Solar
Optical Telescope (SOT) on board \emph{Hinode}. The nonlinear
force-free field (NLFFF) optimization method is adopted to
extrapolate the 3D magnetic fields to higher layers. Three
subregions with different characteristics are selected in the active
region for this study. The results show that the formation heights
in different subregions vary with their different magnetic fields or
velocity patterns. After solving the line blending problem between
the He {\sc \romannumeral 2} 256.32 \AA \ and Si {\sc \romannumeral
10} 256.37 \AA \ lines by the double Gaussian curve fitting, we find
that the transition region lies higher in a strong magnetic area. In
a pre-flare heating area there possibly exist multithermal loops as
implied by comparing the Doppler velocity and the magnetic field on
the solar disk.

\end{abstract}

\keywords{Sun: corona --- Sun: magnetic fields --- Sun: transition
region --- Sun: UV radiation}

\section{Introduction}

The correlation height, defined as the height at which the
correlation coefficient between the Doppler velocity of extreme
ultraviolet (EUV) lines and the magnetic field inclination reaches
its maximum, can be used to represent the formation altitude of
EUV lines \citep{tu2005a, tu2005b}. There are two reasons for
this. On the one hand, the magnetic field controls the plasma
motion in the transition region and corona, where the low $\beta$
condition takes the responsibility. The plasma moves mainly along
the magnetic field lines. Therefore the line-of-sight component of
the velocity should be proportional to the cosine of the angle
between the magnetic field lines and the line of sight, or the
magnetic field inclination. Consequently, the absolute value of
the Doppler velocity has a positive correlation with the value of
$|B_z/B|$, where $B$ is the magnetic field strength and $B_z$ is
the line-of-sight component. Note that the cartesian coordinate
system is adopted here. On the other hand, the EUV line intensity
from an optically thin plasma is proportional to the integration
$\int_{V} G(T)n_e^2 ~\mathrm{d}V$, where $G(T)$ is the
contribution function for plasma with temperature $T$ and $n_e$ is
the electron density \citep{watanabe2007}. Although the volume $V$
has a spatial extention, the contribution function $G(T)$ peaks
sharply at a temperature $T_{\mathrm{max}}$. The line emission is
mostly from the plasma with temperature $T_{\mathrm{max}}$.
Therefore, we can take the height where the temperature reaches
$T_{\mathrm{max}}$ as the EUV line formation altitude.

The formation heights of EUV lines are very important for us to
understand the structure of the solar upper atmosphere, which has
been simulated through 3D MHD numerical modeling
\citep{gudiksen2005} and synthesis of coronal emission lines
\citep{peter2004}. However, it is difficult to deduce the
formation height from observations. The first effort has been made
by \citet{tu2005a} who studied the transition region lines in the
quiet Sun by the correlation method. \citet{tu2005b} further
proposed the funnel model for solar wind origin.
\citet{marsch2006} studied the structure of the solar transition
region in a polar coronal hole, and found that the correlation
heights differ in the open magnetic field region and the closed
one even for the same emission line. \citet{marsch2004,marsch2008}
and \citet{wiegelmann2005} studied the link between plasma flows
and magnetic fields in active regions and in an equatorial coronal
hole, respectively. The emission heights of coronal bright points
derived from the correlation of Fe {\sc \romannumeral 12}
intensity enhancement and the magnetic fields have been discussed
in \citet{tian2007}. With the correlation method, \citet{tian2008}
studied the sizes of transition region structures in coroanl holes
and the adjacent quiet Sun. Moreover, the magnetic field structure
of quiet Sun and coronal holes has been investigated with the help
of a potential field model in \citet{wiegelmann2004b} and compared
with transition region and coronal EUV images.

The correlation method mentioned above needs the combination of
observed EUV lines and 3D magnetic fields, the latter of which are
usually obtained by force-free field extrapolation of the measured
fields in the lower solar atmosphere, such as the photosphere. In
previous papers a linear force-free field (LFFF) model
\citep{wiegelmann2002} or potential field model was adopted.
Recently, a new nonlinear force-free field (NLFFF) optimization
method was proposed by \citet{wheatland2000} and implemented by
\citet{wiegelmann2004}, which has been widely used for the
extrapolation of 3D magnetic fields. The NLFFF extrapolation
method is more accurate for magnetic field reconstruction
especially in active regions \citep{schrijver2006, metcalf2008},
where the magnetic fields are highly sheared and twisted, the
electric current is large, and the force free parameter $\alpha$
usually changes through the space. The potential field and LFFF
model are inadequate for these active regions. A review on this
topic can be found in \citet{wiegelmann2008}.

In this paper we study the correlation heights between the Doppler
velocity of EUV lines and the magnetic field inclination in an
active region with the observations of EUV Imaging Spectrometer
\citep[EIS;][]{culhane2007} and Solar Optical Telescope
\citep[SOT;][]{tsuneta2008,suematsu2008,ichimoto2008,shimizu2008}
on board \emph{Hinode} \citep{kosugi2007}. In particular we adopt
the NLFFF optimization method implemented by
\citet{wiegelmann2004} to extrapolate the 3D magnetic fields.
Three subregions with different characteristics are selected in
the active region for this study. The details of data analysis and
image co-alignment are described in \S~\ref{sec2}. Results are
presented in \S~\ref{sec3}. Discussion and conclusions are made in
\S~\ref{sec4}.

\section{Observations and Data Analysis} \label{sec2}
\subsection{EIS Doppler Velocity Fitting} \label{sec21}

We select six EUV emission lines as shown in Table \ref{tbl1} to
calculate the correlation heights. The six emission lines cover a
relatively large range of formation temperatures, from the
transition region to the corona. They were obtained by the EIS
$1'' \times 256''$ slit scanning from west to east during 19:07 --
23:46 UT on 2006 December 12 in the active region NOAA 10930. The
exposure time is 30 s for each slit. We apply the standard EIS
procedure to calibrate the data and fit the lines with single
Gaussian curve. The Doppler velocities can be calculated by the
line center shifts. When determining the center of a specific
line, two instrumental effects should be taken into account. One
is the slit tilt, because the EIS slit is not perfectly parallel
to the pixel array of the CCD. The other is the orbital variation,
which is caused by thermal effect on the instrument and follows a
sinusoidal behavior. To remove the former, we subtract the Doppler
shift taken early in the mission (regarded as purely due to the
tilt effect) from the current one. To remove the latter, we first
set a rest line center position for each raster which is taken to
be the average of the southern part (a relatively quiet region) of
each slit. Then the time variation of the rest line center
relative to its theoretical position in laboratory is obtained.
After the above steps, we can finally construct the Dopplergrams
for each line.

\begin{figure}
\includegraphics[width=6.2in]{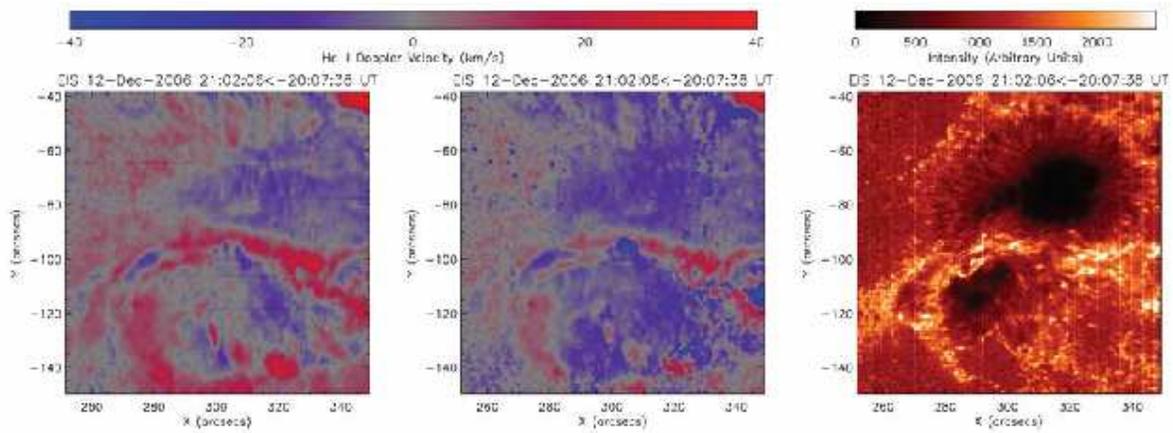}
\caption{EIS He {\sc \romannumeral 2} 256 \AA \ velocity maps
fitted by single Gaussian curve (\emph{left panel}) and double
Gaussian curves (\emph{middle panel}). The velocities that excess
40/$-$40 $\mathrm{km\ s}^{-1}$\ have been set to the top color
(Red/Blue) in order to make the velocity field pattern stand out.
The Ca {\sc \romannumeral 2} H image (\emph{right panel}) shows
that it has similar configurations with the He {\sc \romannumeral
2} 256 \AA \ velocity maps. The Ca {\sc \romannumeral 2} H image
observed by SOT/BFI is created artificially to match the EIS map
in time. See text for details.} \label{fig1}
\end{figure}

The average values of $\chi^2$ in the line profile fitting in some
regions of interest, which will be defined in \S~\ref{sec24}, are
shown in Table \ref{tbl1}. They are relatively small for Fe {\sc
\romannumeral 8}, Fe {\sc \romannumeral 10}, Fe {\sc \romannumeral
13}, and Fe {\sc \romannumeral 15} lines compared to He {\sc
\romannumeral 2} and Fe {\sc \romannumeral 11} lines in all the
regions, since the latter two lines are blended with other lines.
The He {\sc \romannumeral 2} 256.32 {\AA} line is blended with the
Si {\sc \romannumeral 10} 256.37 {\AA} line, especially in active
regions and magnetic loops. Therefore, we use a double Gaussian
curves to fit the two components with only one assumption that the
blueward component corresponds to the He {\sc \romannumeral 2}
line. The Fe {\sc \romannumeral 11} line is a doublet at 188.23
{\AA} and 188.30 {\AA} \citep{young2007a}. We also fit it by
double Gaussian curves assuming that the 188.23 {\AA} component
lies blueward. Note that the double Gaussian curve fitting is
sensitive to the selection of initial values, in particular the
line center wavelengths. Here, we set them to be the rest
wavelengths of each line. The fitting yields seven parameters,
including the amplitudes, widths, and central wavelengths for both
components, and a shared background. The result of double Gaussian
curve fitting for the He {\sc \romannumeral 2} line is shown in
Figure \ref{fig1} and Table \ref{tbl1}. We find that the average
velocities are more blue shifted compared to the result from
single Gaussian curve fitting, as expected. In particular, the
$\chi^2$ values are much smaller in the former case as show in
Table \ref{tbl1}, implying that the fitting goodness is
significantly improved. The result for the Fe {\sc \romannumeral
11} 188.23 {\AA} line is similar to the He {\sc \romannumeral 2}
line.

The non-Gaussian line profiles of EIS in this active region have
been reported by \citet{imada2008}, who fitted the Fe {\sc
\romannumeral 14} 274.20 {\AA} line in the arcades during the
flare. \citet{asai2008} performed the first spectroscopic
observation of an MHD fast mode shock wave during the flare using
double Gaussian curve fitting of the lines. Here, we analyze the
lines before the flare.

\subsection{Magnetic Field Extrapolation} \label{sec22}
\begin{figure}
\includegraphics[width=6.2in]{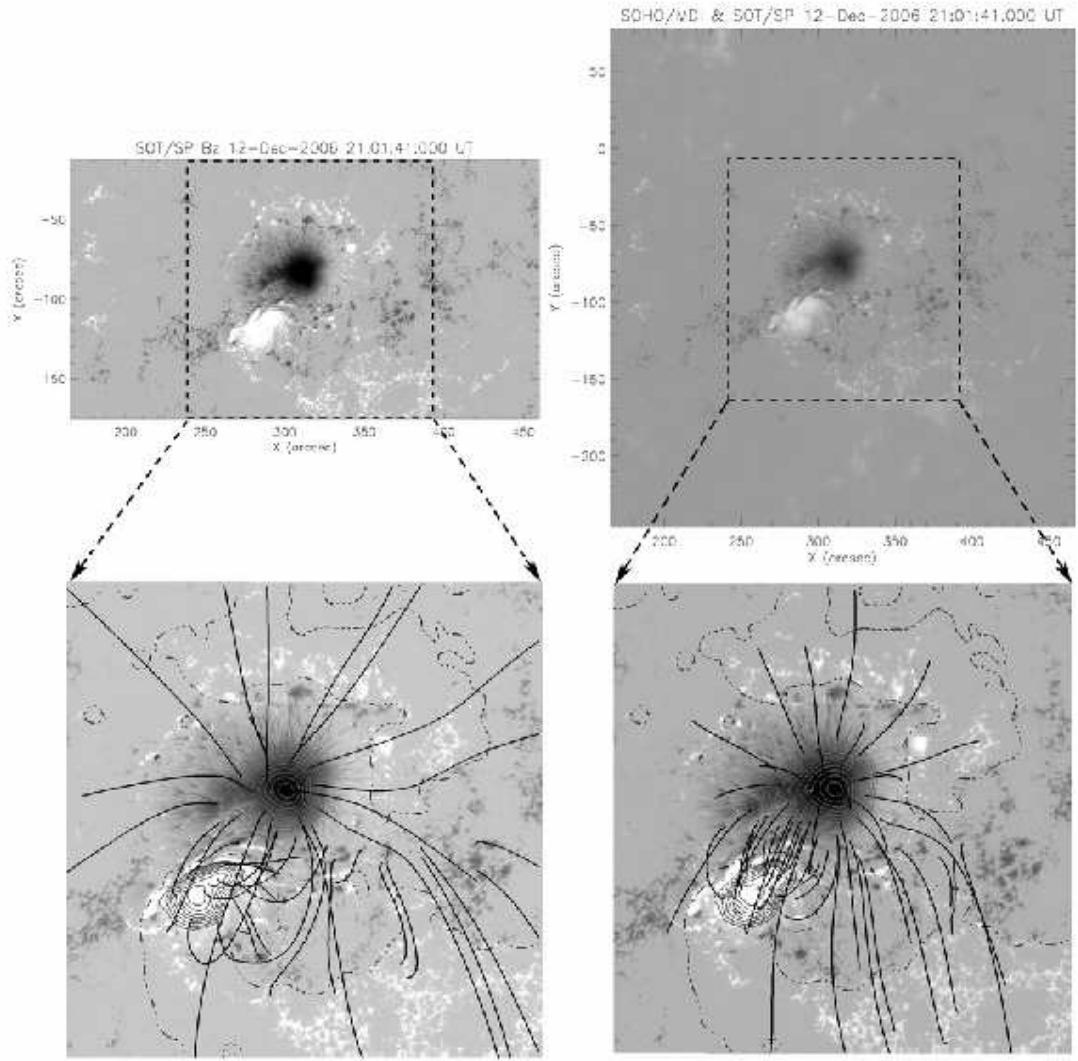}
\caption{\emph{Top left}: The FOV of the magnetic field observed
by SOT/SP from 20:30 to 21:33 UT on 2006 December 12. The dashed
square denotes the FOV in which the NLFFF extrapolation is
performed. \emph{Bottom left}: The 3D magnetic field extrapolated
by the NLFFF model for the dashed square shown in the top left
panel. \emph{Top right}: The FOV of MDI magnetogram for potential
field extrapolation observed at 20:51 UT on 2006 December 12. The
dashed square is the same to that in top left panel. \emph{Bottom
right}: The 3D magnetic field extrapolated by the potential field
model. Note that only the region in the dashed square is shown.}
\label{fig2}
\end{figure}

The active region NOAA 10930 has been studied by many authors
\citep{kubo2007,zhang2007,jing2008,schrijver2008,wang2008}.
\citet{guo2008} extrapolated the 3D magnetic field using the NLFFF
optimization method and studied the configuration, energy release
and shear angle changes in this active region. The photospheric
vector magnetic fields are obtained by fitting the polarized
spectrum observed by SOT/SP (spectropolarimeter), which scanned
the southwestern part of the solar disk from 20:30 to 21:33 UT on
2006 December 12. In the fast map mode, the spatial sampling of
SOT/SP is about $0.3''$. The spectra are fitted by a nonlinear
least-square fitting procedure based on the Milne-Eddington model.
The $180^\circ$ ambiguity of the azimuth angle is removed by
comparing the observed field to the extrapolated LFFF field
\citep{wang1997,wang2001,metcalf2006}. The top left panel in
Figure \ref{fig2} shows the full field of view (FOV) of the
magnetic field observed by SOT/SP. We select half of the FOV and
sample them to $128 \times 128$ grids by $4 \times 4$ binning due
to the limitation in computations. Then the boundary data are
preprocessed by a preprocessing routine developed by
\citet{wiegelmann2006}. The final extrapolation result by the
NLFFF optimization method \citep{wheatland2000,wiegelmann2004} is
shown in the bottom left panel of Figure \ref{fig2}.

We adopt the potential field model to extrapolate the magnetic
field in a larger FOV, because the extrapolation height of SOT/SP
magnetic field is limited to $\sim 110$ Mm and the potential field
model needs only the line-of-sight component of the field, which
can be obtained in the MDI full disk observations. The FOV of MDI
data used for extrapolation is shown in Figure \ref{fig2}
(\emph{top right}), but the data in the area corresponding to the
FOV of the top left panel are replaced by SOT/SP magnetic fields.
This is because that the MDI data suffer from the saturation
problem, which is discussed in detail in a previous paper
\citep{moon2007}. We extrapolate the fields with the Fourier
transform method \citep{aliss1981,gary1989} in this larger FOV
that is four times the FOV used in the NLFFF method. The boundary
data are sampled to $256 \times 256$ grids with spatial resolution
of $\sim 1.2''$. So the extrapolation height is $\sim 220$ Mm. The
3D magnetic field is shown in Figure \ref{fig2} (\emph{bottom
right}). In order to calculate the correlation heights, we have to
align the EUV images with the SOT/SP images. This is done by
correlating the two set of images and finding the offsets between
them.

\subsection{Co-alignment of \emph{Hinode} Observations} \label{sec23}

EIS covers two wavelength bands, i.e. short wavelength (SW) band
(170 -- 211 \AA) and long wavelength (LW) band (246 -- 292 \AA),
which are recorded by two CCDs, respectively. Images on the SW CCD
have an offset of 16 -- $20''$ northward and $2''$ westward
compared to that on the LW CCD \citep{young2007}. We correct for
the offsets in SW images by applying a fixed value of $16''$ in
the north-south direction and $2''$ in the east-west direction. We
also need to align the SW and the LW images with the magnetograms.
The main idea is to align all the images by EIS and SOT with that
by EIT \citep[EUV Imaging Telescope;][]{dela1995} and MDI
\citep[Michelson Doppler Imager;][]{scherrer1995} on \emph{SOHO}
(\emph{Solar and Heliospheric Observatory}). The full disk images
by EIT and MDI enable us to align them by comparing the limbs. The
accuracy is close to their spatial resolutions. Since the EIS and
SOT observations are within partial disk, we cannot get the
absolute coordinates of the solar disk center. However, after
being aligned with the MDI images, all the coordinates are
referred to the true solar disk center.

The SOT/SP magnetogram is aligned with the MDI magnetogram as
follows. First of all, the two images are interpolated to the same
resolution, always to the better one. Then, we select two slit
images to do the correlation because the MDI magnetogram was
observed during the scanning time of the SOT/SP one, i.e. 20:30 --
21:33 UT. The slit image of SOT/SP that has the closest time to
the MDI observation is correlated with the MDI slit images in
different positions. Finally, the offset is found for which the
two slit images have the largest correlation coefficient. The
uncertainty of the co-alignment is about $2.0''$.

The EIS Fe {\sc \romannumeral 12} 195~\AA \ image is aligned with
the EIT Fe {\sc \romannumeral 12} 195~\AA \ image with the
correlation method, too. Then the EIT 195~\AA \ image is aligned
with the MDI magnetogram by comparing the limbs. Reading the
coordinates of the east, west, south, and north poles on the solar
limb in the MDI magnetogram, we calculate the coordinates of the
disk center and find that the coordinates provided by MDI are
correct. But the EIT 195~\AA \ image taken at 23:49 UT on the same
day, which is the closest to the EIS observation, shifts $4.0''$
westward and $6.0''$ northward in the coordinates provided by the
standard software. The uncertainty of the shift is $\sim 2.6''$,
which is the resolution of EIT images. Finally, The offset between
the EIS Fe {\sc \romannumeral 12} 195~\AA \ image and the MDI
image is found. The total uncertainty is $\sim 5.2''$.

The EIS He {\sc \romannumeral 2} 256~\AA \ velocity map is aligned
with the SOT/BFI (Broadband Filter Imager) Ca {\sc \romannumeral
2} H image, the SOT/SP and MDI magnetograms consecutively. We find
that the red shift region of the He {\sc \romannumeral 2} velocity
map has similar configurations with the heating region of the Ca
{\sc \romannumeral 2} H image as shown in Figure \ref{fig1}. Note
that the Ca {\sc \romannumeral 2} H image was observed every 2
min, but the EIS map was constructed from raster scanning during
about 4.5 hours. We need to reconstruct a Ca {\sc \romannumeral 2}
H image in the exactly same manner as the EIS map. To do so, we
cut the Ca {\sc \romannumeral 2} H images at each time into slices
along the slit. We pick out the slice with the same position and
time as each slit in EIS scanning observations. Then we adopt a
feature identification method to align them, i.e. select points
with similar features alternately from the two images. Usually we
select five points from each image to calculate an offset and
repeat it 10 times to estimate the uncertainties. The average
offsets in x and y directions are $4.3''$ and $-7.8''$, with
errors of $1.7''$ and $1.2''$, respectively. Next, the Ca {\sc
\romannumeral 2} H image is aligned with the SOT/SP magnetogram
with the identification method, too. The offset in x direction is
$-0.6''$ with an error of $0.4''$, and that in y direction is
$-7.1''$ with an error of $0.4''$. Because the SOP/SP magnetogram
has already been aligned with the MDI magnetogram, the EIS He {\sc
\romannumeral 2} velocity map can be aligned with the latter with
total uncertainties of $4.1''$ and $3.6''$ along x and y
directions, respectively.

\begin{figure}
\includegraphics[width=3.5in]{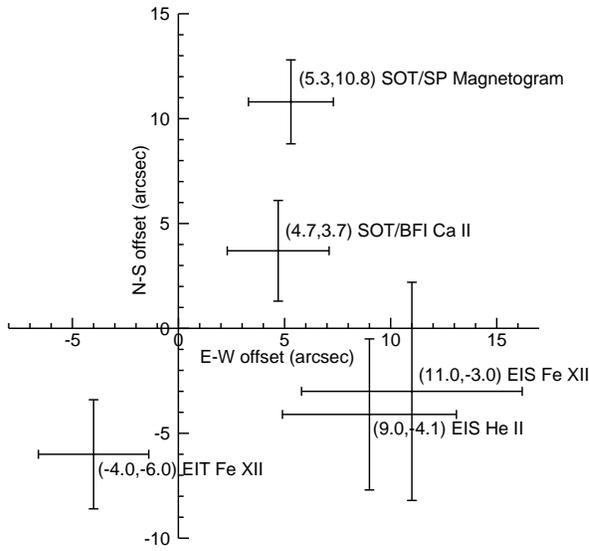}
\caption{Offsets of images observed by different instruments with
respect to MDI. Plus values in the parentheses refer to that image
shifts to east or south, while negative values to west or north.
The error bars are also plotted, both in the east-west direction
and the north-south direction.} \label{fit3}
\end{figure}

We plot the offsets of all images with respect to the MDI image in
Figure \ref{fit3}. If we adopt the offsets between SW and LW
images given by \citet{young2007}, we only have to align one of
them with MDI magnetograms. Then the other can be aligned
automatically. But here we align both with independent methods to
check the offsets derived by us and by them. The LW and SW images
have the same offsets with respect to MDI magnetogram within the
uncertainties, which indicates our results are coincident with
theirs.

\subsection{Correlation Heights and Error Estimation} \label{sec24}

\begin{figure}
\includegraphics[width=6.2in]{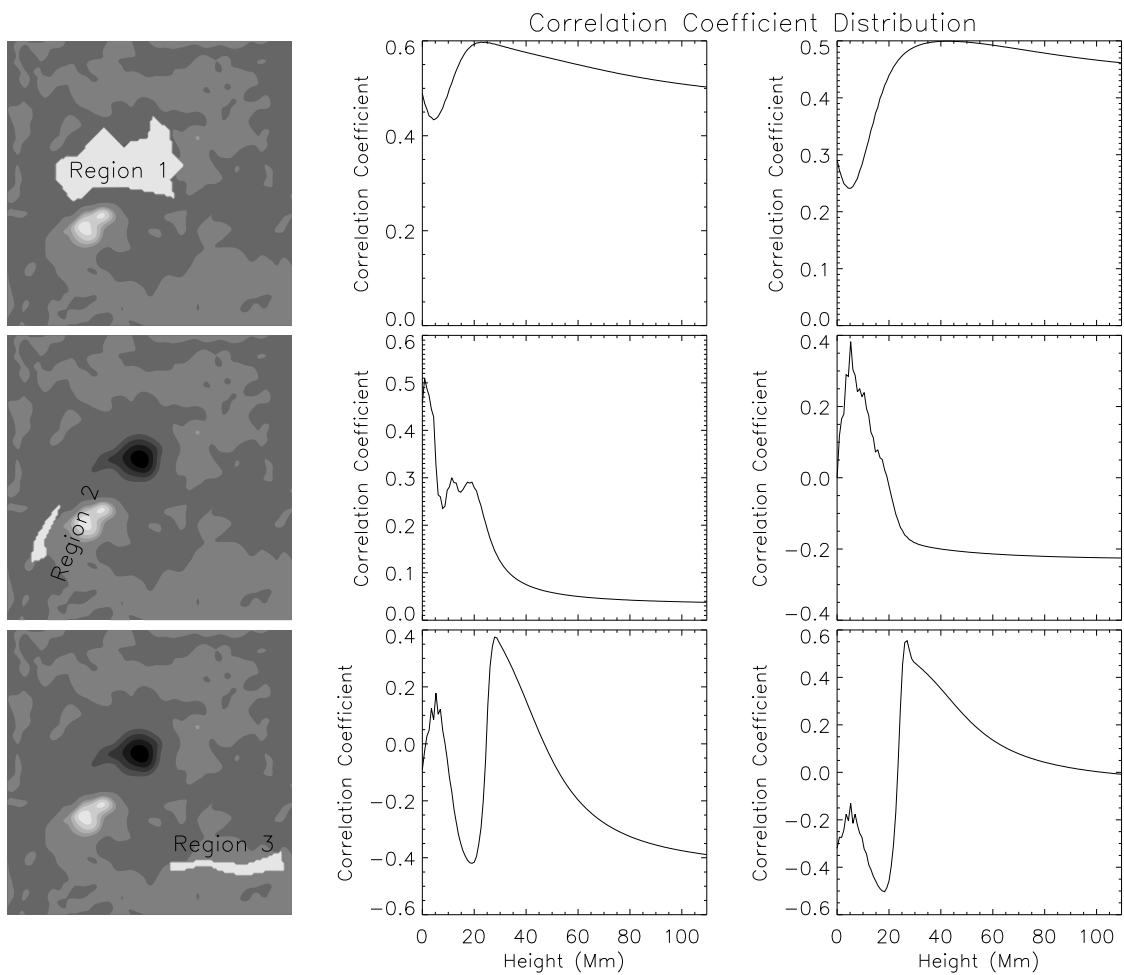}
\caption{Height distributions of the correlation coefficient for
the He {\sc \romannumeral 2} 256 \AA \ line in different
subregions of the active region NOAA 10930. The white polygons
labelled with Regions 1 -- 3 (\emph{left column}) show the areas
selected to calculate the correlation coefficients. The
distributions in middle and right columns are calculated by
correlating the magnetic field inclination with the Doppler
velocity, which is obtained by single and double Gaussian curve
fitting, respectively. The distributions in top to bottom rows are
obtained in Regions 1 -- 3, respectively.} \label{fig4}
\end{figure}

We select different regions to calculate the correlation
coefficients that are plotted in Figure \ref{fig4}. Different from
quiet Sun regions, the environment in active region is quite
inhomogeneous. It is necessary to select different subregions to
calculate the correlation heights. The high spatial resolution
observations of Hinode provide us the possibility to do so. The
subregions are selected according to spatial patterns of the
magnetic fields and velocity fields; so their shapes are not
necessarily regular. Region 1 mainly concentrates on the negative
magnetic polarity region and the blue shift area of He {\sc
\romannumeral 2} 256 \AA. Region 2 is on the east side of the
polarity inversion line corresponding to the foot points of some
lower loops and the red shift area of He {\sc \romannumeral 2} 256
\AA. Region 3 is within the projection area of the higher closed
magnetic loops. The correlation coefficient distributions are
calculated by correlating the magnetic field inclination with the
Doppler velocity, which is obtained by single (\emph{middle
column}) and double (\emph{right column}) Gaussian curve fitting
in Figure \ref{fig4}, respectively. The height is less than 5 Mm
in Region 2, but more than 20 Mm in Regions 1 and 3. For different
lines listed in Table \ref{tbl1}, the correlation heights can be
calculated with the same method.

There are errors in the image co-alignment, which also cause errors
in calculations of the correlation height. As mentioned in
\S~\ref{sec23}, there is an offset between SW and LW EIS images.
Moreover, different lines in the same band have internal pointing
shifts, which are $0.5''$ in \emph{x} direction and $1''$ in
\emph{y} direction for SW and $3''$ in \emph{x} direction and $4''$
in \emph{y} direction for LW as shown in a preliminary research by
Deb\footnote{http://msslxr.mssl.ucl.ac.uk:8080/eiswiki/Wiki.jsp?page=FitsPointing}.
Taking into account these facts, the errors in the image
co-alignment are $7.7''$ in \emph{x} direction and $8.2''$ in
\emph{y} direction for SW except for Fe {\sc \romannumeral 12}
195~\AA. They are $7.1''$ in \emph{x} direction and $7.6''$ in
\emph{y} direction for LW except for He {\sc \romannumeral 2}
256~\AA. The errors of the correlation height are estimated by the
following way. First, we artificially shift the EIS velocity map
relative to the magnetic field $B_z$ image within the range of
co-alignment errors. Then, the correlation heights are calculated
for each case (here, we select $1'' \times 1''$ grids). Finally, we
get a grid of correlation heights with different correlation
coefficients. The final correlation height for one specific line is
estimated as the average of the heights with correlation
coefficients greater than 0.07, and the error is the standard
deviation of these heights.

\section{Results} \label{sec3}

The average formation height of He {\sc \romannumeral 2} 256 \AA \
in region 2 is lower than that in Regions 1 and 3. In comparison,
the mean magnetic fields in Region 2 and 3 are one order of
magnitude weaker than that in Region 1. This implies that stronger
magnetic fields correspond to higher formation heights of EUV
lines. For Regions 2 and 3, the mean magnetic fields are almost
the same, but the Doppler velocity patterns are different, which
show mainly red shifts and blue shifts, respectively. This means
that down flows tend to make the formation height lower. In
summary, the correlation heights are different in various regions
even for the same line He {\sc \romannumeral 2} 256 \AA \ because
of different physical conditions.

\begin{figure}
\includegraphics[width=6.2in]{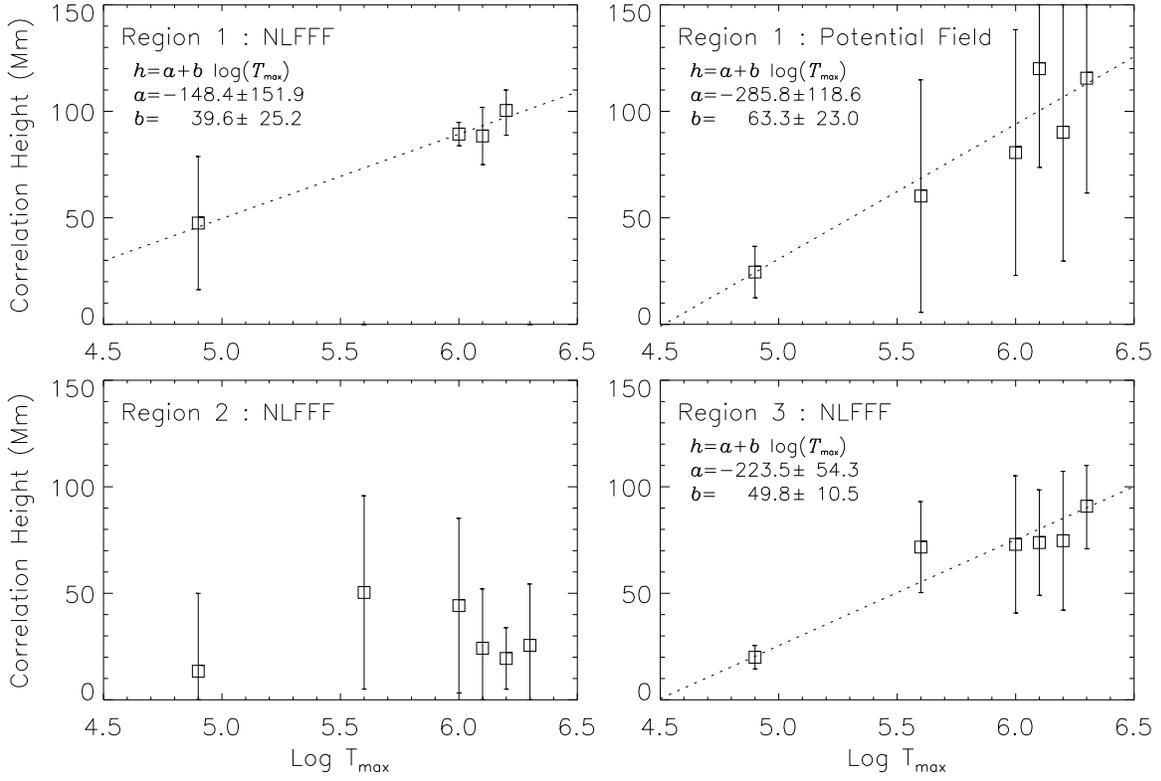}
\caption{Correlation heights versus log $T_\mathrm{max}$ for Regions
1 -- 3. The magnetic field is extrapolated with the NLFFF model
except for the top right panel, in which the potential field model
is used. The dotted lines are the linear fitting of the height --
log $T_\mathrm{max}$ relationship.} \label{fig5}
\end{figure}

The correlation height versus log $T_\mathrm{max}$ is plotted in
Figure \ref{fig5}, where the EUV line formation temperature,
$T_\mathrm{max}$, is listed in Table \ref{tbl1} and the errors of
formation heights are derived using the method mentioned in
\S~\ref{sec24}. In Region 1, shown in Figure \ref{fig5} (\emph{top
left}), there are no points for Fe {\sc \romannumeral 8} and Fe
{\sc \romannumeral 15}, since we cannot find the heights with
correlation coefficients greater than 0.07. A possible reason is
that the real formation height of Fe {\sc \romannumeral 15}
exceeds the NLFFF extrapolation domain, which is about 110 Mm in
height. For Fe {\sc \romannumeral8} we still do not know the
reason. The correlation heights for all the EUV lines can be got
in Region 1 by the potential field extrapolation in Figure
\ref{fig5} (\emph{top right}), which can reach about 220 Mm in
height. For the regions studied here, the heights for the same
line calculated with different extrapolation methods differ
slightly within the errors. Also, the heights calculated by the
potential field give a similar height versus log $T_\mathrm{max}$
relationship compared to that calculated by NLFFF. However, the
NLFFF gives less uncertainties for the correlation heights and
makes the results more confirmative.

Region 2, shown in Figure \ref{fig5} (\emph{bottom left}), is a
pre-flare heating area. Massive downflows can be seen in the EIS
velocity map (Figure \ref{fig1}). All the EUV lines are formed
within a smaller range of heights. The temperature structure is
complicated in this area. There is no clear height versus log
$T_\mathrm{max}$ relationship in Region 2, while there are positive
relationships in Region 1 and Region 3 (\emph{bottom right}) as
shown in Figure \ref{fig5} . But the temperature in Region 3 rises
more quickly. For example, log $T_\mathrm{max}=6.3$ corresponds a
height of $\sim$ 90 Mm in Region 3, compared to $\gtrsim$ 100 Mm in
Region 1. This is probably due to the higher magnetic field strength
in Region 1, which seems to suppress the temperature increase with
height. Therefore, the EUV line formation height is affected by not
only the magnetic topology but also the field strength.

We also plot the linear fitting of the height -- log
$T_\mathrm{max}$ relationship for Regions 1 and 3 in Figure
\ref{fig5}. These curves provide a quantitative description of how
the temperature varies with height in the corona.

\section{Discussion and Conclusions} \label{sec4}

The EUV lines in an active region are formed at different heights
according to the physical conditions of the magnetized atmosphere
in situ, including the velocity, density, temperature, and
magnetic field. In this paper we examine some of these factors
with the correlation height method by aligning images observed by
SOT and EIS. In particular, we find that the transition region,
where He {\sc \romannumeral 2} 256 \AA \ is formed, lies higher in
areas with strong magnetic fields.

Results from Region 2 show that the EUV emission lines are formed
within a relatively narrow height range in the pre-flare heating
loops. A possible scenario is that this region contains coronal
loops that are multithermal. \citet{moe1998} have shown that some
active region loops contain mixtures of plasma of different
temperatures. Multithermal plasma has also been found in the
transequatorial loops \citep{brosius2006}. All these researches
are based on the analysis of the co-spatial emission lines on the
solar limb. While we find a similar result by analyzing the lines
on the solar disk and correlating the Doppler velocity and
magnetic field.

\acknowledgments

We thank the referee for constructive comments that helped to
improve the paper. Hinode is a Japanese mission developed and
launched by ISAS/JAXA, with NAOJ as domestic partner and NASA and
STFC (UK) as international partners. It is operated by these
agencies in co-operation with ESA and NSC (Norway). Y. Guo, M. D.
Ding, \& M. Jin were supported by National Naturel Science
Foundation of China (NSFC) under grants 10878002, 10333040, and
10673004 and by NKBRSF under grant 2006CB806302. T. Wiegelmann was
supported by DLR-grant 50 OC 0501.

\clearpage
\begin{table}
\caption{EIS emission lines, average Doppler velocities, and average
values of $\chi^2$~for the line profile fitting in three selected
regions.}\label{tbl1}
\begin{tabular}{c c c r r r r r r}
\\ \hline \hline
Ion & Wavelength & log $T_{\mathrm{max}}$ & \multicolumn{3}{c}{Average Doppler Velocity${}^\mathrm{a}$} & \multicolumn{3}{c}{Average $\chi^2$${}^\mathrm{a}$} \\
 & (\AA) &  & \multicolumn{3}{c}{(km s${}^{-1}$)} & \multicolumn{3}{c}{}\\
\hline
He {\sc \romannumeral 2}${}^\mathrm{b}$........ & 256.32 & 4.9 & $\quad$-9.8 & $\: \: \quad$4.3 & -0.3$\quad$ & $\qquad$1.6 & $\; \; \quad$5.0 & 3.3$\qquad$ \\
He {\sc \romannumeral 2}${}^\mathrm{c}$........ & & & -12.9 & 1.7 & -2.7$\quad$ & 0.5 & 0.4 & 0.5$\qquad$ \\
Fe {\sc \romannumeral 8}...... & 185.21 & 5.6 & -21.9 & 1.6 & -5.2$\quad$ & 0.3 & 0.3 & 0.3$\qquad$ \\
Fe {\sc \romannumeral 10}......... & 184.54 & 6.0 & -27.5 & 1.2 & -13.8$\quad$ & 0.4 & 0.7 & 0.6$\qquad$ \\
Fe {\sc \romannumeral 11}${}^\mathrm{b}$....... & 188.23 & 6.1 & -15.4 & -1.0 & -9.9$\quad$ & 2.5 & 12.4 & 10.1$\qquad$ \\
Fe {\sc \romannumeral 11}${}^\mathrm{c}$....... & & & -25.3 & -9.7
& -14.4$\quad$ & 0.5 & 0.3 & 0.5$\qquad$ \\
Fe {\sc \romannumeral 13}...... & 202.04 & 6.2 & -24.3 & -4.0 & -10.9$\quad$ & 0.3 & 0.4 & 0.4$\qquad$ \\
Fe {\sc \romannumeral 15}....... & 284.16 & 6.3 & -11.8 & -5.3 & -1.7$\quad$ & 0.7 & 2.0 & 1.2$\qquad$ \\
\hline
\multicolumn{9}{l}{} \\

\multicolumn{9}{l}{${}^\mathrm{a}$~The values in the three columns
are calculated for Regions 1 -- 3, which are defined in Figure
\ref{fig4}.} \\

\multicolumn{9}{l}{${}^\mathrm{b}$~The average velocities and
$\chi^2$ for this line are calculated by single Gaussian curve
fitting.} \\

\multicolumn{9}{l}{${}^\mathrm{c}$~The average velocities and
$\chi^2$ for this line are calculated by double Gaussian curve
fitting.}
\end{tabular}
\end{table}

\clearpage

\end{document}